\documentclass[12pt,floatfix,superscriptaddress,aps,prd,preprint,nofootinbib]{revtex4}
\usepackage{amsmath,amssymb,amsbsy,amsfonts,amsopn,amstext}
\usepackage{graphicx}
\usepackage{xcolor,color}
\usepackage{slashed}
\usepackage{esint,ulem}
\usepackage[dvips]{epsfig}
\usepackage[dvips]{graphicx}
\usepackage{units}
\usepackage{textcomp}
\newcommand{\beq}{\begin{equation}}
\newcommand{\eeq}{\end{equation}}
\newcommand{\bea}{\begin{eqnarray}}
\newcommand{\eea}{\end{eqnarray}}

\begin{document}
\title{New regular 2+1 black hole solutions from bilocal gravity}
\author{H. R. Christiansen\footnote{E-mail: hugo.christiansen@ifce.edu.br}} \affiliation{Instituto Federal de Ci\^encias, Educa\c{c}\~ao e Tecnologia (IFCE) 62042-030 Sobral and  61940-750 Maranguape, CE,  Brazil.}
\author{Milko Estrada \footnote{E-mail: milko.estrada@gmail.com}}\affiliation{Facultad de Ingenier\'ia, Ciencia y Tecnolog\'ia, Universidad Bernardo O'Higgins, Santiago, Chile.}
\author{M. S. Cunha\footnote{E-mail: marcony.cunha@uece.br}}\affiliation{Universidade Estadual do Cear\'a, Centro de Ci\^encias e Tecnologia, 60714-903, Fortaleza, CE, Brazil.}
\author{J. Furtado\footnote{E-mail: job.furtado@ufca.edu.br}}\affiliation{Universidade Federal do Cariri, Centro de Ci\^encias e Tecnologia, 63048-080, Juazeiro do Norte, CE, Brazil.}
\author{C. R. Muniz\footnote{Corresponding author. E-mail: celio.muniz@uece.br}}\affiliation{Universidade Estadual do Cear\'a, Faculdade de Educa\c c\~ao, Ci\^encias e Letras de Iguatu, 63500-000, Iguatu, CE, Brazil.}

\begin{abstract}
We obtain new regular black hole solutions for an action in 2+1 dimensions with bilocal Ricci scalar and negative cosmological constant. Besides their connection to the cosmological constant, these solutions depend on a fundamental length due to their non-local nature. The effective profile densities that result from the non-local geometries have quasi-localized mass/energy since they are finite at the origin and their integration in all space is convergent.  The black holes obtained are free of singularities and present one, two, or none horizons depending on the values of the involved parameters. The new solutions can have either an AdS, dS, or even a flat core. In the case of a de-Sitter core, it could represent a repulsive force coming from quantum effects. Although the resulting (effective) cosmological constant is positive near the origin, the classical (naked) counterpart is still negative thus precluding a cosmological horizon. We investigate the energy conditions of the effective source and determine the region where exotic energy should be found. Thermodynamic quantities are also computed. On the one hand, Gibbs's potential shows that both solutions are globally unstable, as in the BTZ case. On the other, we show that for small values of the horizon radius the Hawking temperature is negatively divergent but a finite size remnant can be defined where $T_H$ crosses zero. At this point, the heat capacity sign changes from negative to positive, indicating that the black holes are locally stable while irradiating. Thus, such a quantity, along with $T_H$, presents crucial differences with the BTZ black hole for small horizon radii where quantum effects become relevant. Finally, we analyze the bilocal black hole geodesics and find stable circular orbits for massless and massive particles, another feature absent in the BTZ case.\\
~
\\

\end{abstract}
%
%\pacs{72.80.Le, 72.15.Nj, 11.30.Rd} esses números não são da área de cosmologia e gravitação
%
\maketitle
\section{Introduction}

Nonlocality is a well-established subject of research in the pursuit of regular complete theories of fundamental interactions, particularly in the ultra-high energy regime \cite{nonlocal}.
For a long time, an approach based on nonlocality has been disregarded due to the initial success of regularization procedures. However, when the case is gravity, the usual renormalization scheme runs into serious difficulties and nonlocality might be the unique way out \cite{deserVN74}.

Since it is expected that space-time changes its nature when probed at distances of the order of the Planck length, the usual description in terms of smooth differential manifolds would be just the low energy limit for a highly excited, granular space-time. The associated loss of resolution imposes a fundamental length cutoff to treat the inappropriate short-distance behavior of the theory. This is at the root of every approach to a quantum theory of gravity, such as string theory, loop quantum gravity, generalized uncertainty principle, and others.

It is worth noting that the equations of motion emerging from non-local classical actions are in general acausal. This is an important reason to stress that the fundamental functional we will use is local. Note that the variation of the effective action does not give the equations of motion of the fields but rather the equations of motion of their vacuum expectation values. In general, nonlocalities result from the quantum fluctuations of the lightest fields in the theory, namely the graviton. Since one is not integrating out any field, the quantum effective action has the same domain as the fundamental theory and is not just a low-energy theory.

Additionally, several studies have considered a reliable alternative to the singularity formation hypothesis
of the current theories of quantum gravity. A black hole would be a state for which the energy density eventually reaches a (finite) maximum of the order of the Planck scale; see e.g. \cite{Spallucci:2017aod}. So, while a central singularity is avoided a regular black hole is obtained. It is worth mentioning that the equations of motion in this context are classically solved. In \cite{Rovelli:2014cta} the same topic is discussed speculating that a quantum phase is formed when the gravitational attraction is balanced with a very large quantum pressure. These kinds of models are dubbed Planck stars. Ref. \cite{DeLorenzo:2014pta} argues that the metric of Planck stars has a de-Sitter core whose effective cosmological constant causes a repulsive force resulting from to quantum effects.
Thus, it is of physical interest to study nonlocal scenarios of gravity that lead to regular black hole solutions with a finite energy density of the order of the Planck scales near the origin.
%
%\footnote{Nonlocal effects are well understood in the ultraviolet regime but not so much in the infrared, where they could give rise to cosmological issues. Cosmological models are usually developed starting from a classical action and the background evolution with the cosmological perturbations derived from it. However, the relevant action is the quantum effective action which is unavoidably not local in principle.  The techniques for computing these nonlocal terms are well understood at high energies but at low energies the situation is much less clear. See eg. \cite{Belgacem} for a model which aims validity at energies well below the Planck scale.}
%we must distinguish between the in-out
%and the in-in expectation values. The standard Feynman path integral gives in-out vacuum
% exp.val. These obey a-causal equations involving the Feynman propagator.
%There is nothing wrong with it, since in-out elements are
%not physically observable and only useful as intermediate steps incomputation of eg. scattering cross section.
%On the other hand, in-in elements are observables.

%%%%%%%

In three-dimensional spacetime, which is our framework in this paper, one has to take into account some peculiarities. Interestingly, there is a close connection of 2+1 gravity with pure Chern-Simons gauge theory \cite{achucaro86, witten88}, as well as with quasi topological electromagnetism \cite{cisterna2020}, which has allowed gauge field arena to study gravity in 3D. Nothing so extraordinary is found in 3+1 dimensions. Furthermore, the existence of gravitational waves with nonlinear interactions in 3+1 dimensions means that one has no chance of an exact solution of any system along with quantum gravity. Moreover, quantum gravity with $\Lambda>0$ (as it is in our world) could possibly not exist nonperturbatively in any dimension \cite{lambda>0}.
On the other hand, in three dimensions there are different shortcomings. The 2+1 Einstein-Hilbert theory at the classical level is trivial, in the sense that there are no gravitational waves, and any two solutions are locally equivalent. The (2+1)-dimensional vacuum  has no local degrees of freedom \cite{leutwyler}.
Hence, black holes were unexpected since a vacuum solution in 2+1 dimensions is necessarily flat. The reason is that in 2+1 the Weyl tensor is zero and the Ricci tensor vanishes due to the Einstein field equations. Consequently, the full Riemann tensor is trivial and no black hole (BH) solution with event horizons exists.
Finally, for $\Lambda\geq0$, there is no non-degenerate (nor degenerate) apparent horizon \cite{ida2000}.
Thus, three-dimensional $\Lambda<0$ would be the best opportunity for a model with quantum singularities. These are the honored BTZ black holes \cite{BTZ}.
Their existence makes 3D gravity exciting and one may guess whether it could be tractable quantum mechanically.
With a strictly negative cosmological constant, a 2+1 BH solution emerges presenting similar properties to the (3+1)-dimensional Schwarzschild and Kerr black holes. Remarkably, it admits a no-hair theorem thus characterizing the solution by its ADM mass, angular momentum, and charge.
It has also the same thermodynamic properties such as BH entropy, which obeys a kind of Bekenstein bound where the surface area is replaced by the BTZ black hole's disk.
Likewise, a rotating BTZ BH contains an inner and an outer horizon, analogous to an ergosphere in a Kerr BH.
Since (2+1)-dimensional gravity has no Newtonian limit, the BTZ BH might not be the final state of a gravitational collapse. However one can calculate the BTZ energy-momentum tensor as in 3+1 black holes enabling an eventual collapse \cite{carlip1995}.
To recast our brief overview, in a spatially compact manifold, 2+1 general relativity becomes a topological field theory \cite{martinec,achucaro86} with only a few non-propagating degrees of freedom. There seems to be no place for quantum states to account for BH thermodynamics.
In a non-compact manifold, however, the theory acquires a new set of dynamical degrees of freedom described by an $SL(2,R) \times SL(2,R)$ WZW action or a Liouville action \cite{noncompact}. Whether we can count these degrees of freedom to obtain a microscopic explanation of the entropy of the BTZ BH is a very difficult question not yet completely answered.
BH thermodynamics has been assumed to be the consequence of statistical mechanics of underlying quantum gravitational states although the detailed nature of these states has remained unsolved.
For our aim here, the most important characteristic of the BTZ BH is that it has thermodynamic properties closely analogous to those of realistic
four-dimensional ones. Most significantly, the BTZ BH radiates at a Hawking temperature and has an entropy equal to a quarter of its event horizon area. These features can be obtained in all the usual procedures, a kind o unexpected universality \cite{universality}. In this direction, the Cardy formula for the classical algebra charges yields an entropy agreeing precisely with the Bekenstein-Hawking entropy \cite{cardy}.
Such notable results in a low dimensional setting strongly suggest that BH statistical mechanics in higher dimensions can be successfully investigated within (2+1)-dimensional gravity.
In this paper, in order to avoid the difficulties inherent to the nonlocal character of each specific formulation we will adopt an alternative route to cope with the essential ultraviolet features of gravity by treating nonlocality as coming from the matter sector. We will follow the approach of \cite{moffat gravity,moffat etal2011} which will be explained in the next section. Also, and related to the proposals raised in \cite{Spallucci:2017aod,Rovelli:2014cta,DeLorenzo:2014pta}, we will consider scenarios with negative cosmological and a de-Sitter core in 2+1 dimensions.
%%
%As a final comment, let us mention that is not yet clear whether the quantum states of the black hole can be obtained purely within gravity. It has been argued that for this purpose general relativity must be considered as an effective field theory which cannot distinguish among different conformal field theory states with the same expectation values of the stress-energy tensor. As such, only a more comprehensive microscopic theory (string theory or a dual gauge theory) can describe the true underlying degrees of freedom (see e.g. \cite{martinec98}).
%authors have tried to obtain the BTZ black hole entropy
%by counting states in particular conformal field theories that are, arguably, induced from pure (2+1)-dimensional gravity.
%%
%%%%%%%%
%%
%In order to avoid the difficulties inherent to the nonlocal character of some specific formulations, in this paper we will adopt an alternative route to cope with the essential ultraviolet features of gravity by treating nonlocality as coming from the matter sector. We will follow the approach o Moffat and co-workers \cite{moffat gravity,moffat etal2011} which will be explained in the next section.
%
%
The paper is organized as follows: In section II we analyze the 2+1 bilocal BH solutions and the effect of nonlocality on the gravitational field. In section III we analyze the energy conditions for the source of the modified black hole solutions. Section IV is dedicated to analyzing the Hawking temperature together with black hole volume, heat capacity at constant pressure, and Gibbs free energy. In Section V, we study the circular geodesics of massless and massive particles. In the last section, we draw our conclusions.
\section{Bilocal black hole solutions in 2+1}
The standard procedure to achieve nonlocal effects in a field action consists in considering an infinite order
of derivative terms. The first nonlocal gravity models appearing in the literature added new terms proportional
to $R {F}(\square^{-1}R)$ to the ordinary Ricci scalar term in the Einstein-Hilbert action \cite{nonlocalgrav}.
Among several possibilities, in this paper we will study the effect of
nonlocality on the gravitational field by assuming an action as proposed in \cite{moffat gravity,moffat etal2011,NLAction}, with a recent extension for black strings in \cite{Hugo1}.
The problem with quantum gravity based on a local field theory and a point-like graviton
is not only its nonrenormalizability but also that interactions with matter are not renormalizable at any loop order.
In order to handle this situation, Moffat developed a strategy based on his
previous work in the electroweak quantum field theory
consisting in modifying the vertex couplings by means of entire functions \cite{moffat qft}.
Similarly, in quantum gravity, the coupling constant is minimally modified by
$\sqrt{G}= \mathcal{F}\sqrt{G_N}$ where
$\mathcal{F}$ has no poles in the finite complex plane which could correspond to physical particles in the spectrum.
Any singularity would be just at infinity so the theory
can be UV complete and satisfy unitarity to all orders.
The 3+1 dimensional graviton-graviton and graviton-matter interactions are nonlocal
due to the nonlocal holographic functions at all the vertices in the Feynman graphs  \cite{moffat etal2011}.
In this approach, it is not the pure Einstein-Hilbert action that is modified.
It is the matter action $S_M$ that incorporates the nonlocal coupling to gravity.
The generalized energy-momentum tensor
$-\frac 2{\sqrt g} \frac{\delta S_M}{\delta g^{\mu\nu}}=S_{\mu\nu} $
is now given by
\beq S_{\mu\nu} = \mathcal{F}^2(\Box/2\Lambda_G^2)T_{\mu\nu}\label{ST}\eeq
where $\Box$ is the generally covariant d'Alambertian, $T_{\mu\nu}$
is the stress-energy-momentum tensor in the standard local coupling, and
$\Lambda_G$ is a fundamental gravitational energy scale.
The generalized Einstein field equations are
\beq G_{\mu\nu} =8\pi G_N  S_{\mu\nu} \label{einstein}\eeq
where, as usual,  $G_{\mu\nu}=R_{\mu\nu}-\frac12g_{\mu\nu}R$.
The nonlocal differential operator thus produces a smeared energy-momentum tensor.
Therefore, Eq. \eqref{einstein} describes gravity coupled to a generalized matter source term which includes full nonlocality.
Since $\mathcal{F}$ is invertible, Eq. (\ref{einstein}) can be written as
\beq
\mathcal{F}^{-2}(\Box/2\Lambda_G^2)G_{\mu\nu} =8\pi G_N  T_{\mu\nu}. \label{einstein2}
\eeq
Now, there are two different vacuum states, given by $T_{\mu\nu}=0$ or $S_{\mu\nu}=0$.
Written in this way, Eq. \eqref{einstein2} recasts the same meaning but instead apparently modifies the Ricci tensor.
This is precisely the strategy adopted: generalizing $T_{\mu\nu}$ by
applying the non-local operator as in Eq. (\ref{ST}).
%
%\newline
%$\dots$
%{\color{red}ver se $\mathcal{A}=\mathcal{F}^{-1}$}
%\newline
%
In 2+1 dimensions the modified gravitational action is
\begin{equation}
S_{grav}=\frac{1}{2\kappa}\int d^3 x\sqrt{g}( \mathcal{R}(x)-\Lambda)
\end{equation}
where in place of the Ricci scalar $R(x)$ we have
\begin{equation}
\mathcal{R}(x)=\int d^3y  \sqrt{g}\mathcal{F}^{-2}(x-y)R(y).
\end{equation}
Here $\mathcal{F}^{-2}(x-y)=\mathcal{F}^{-2}(\Box_x)\delta^3(x-y)$ is a bilocal operator and $\Box_x$ is the generally covariant d'Alambertian.
The modified Einstein's equations become
\begin{equation}\label{EME}
G_{\mu\nu}+g_{\mu\nu}\Lambda=\kappa\mathcal{T}_{\mu\nu},
\end{equation}
where $\mathcal{T}_{\mu\nu}=\mathcal{F}^{2}(\Box_x)T_{\mu\nu}$ and $\kappa=8\pi G_N/c^4$ is the Einstein gravitational constant.
We will consider a point static source so that $T_{0}^{0}=-\rho(x)=-M\delta^2(x)$ and
\begin{equation}
\mathcal{F}^{2}(\Box_x)\delta^2(x)=(2\pi)^{-2}\int d^2p\, \mathcal{F}^{2}(-\nabla^2)e^{i{\bf p\cdot x}}=(2\pi)^{-2}\int d^2p\, \mathcal{F}^{2}(p^2)e^{i{\bf p\cdot x}}.
\end{equation}
 The general strategy of theories of this kind is to find a suitable form of $\mathcal{F}^{-2}(\Box_x)$ so that the
Euclidean momentum space function $\mathcal{F}^{2}(p^2)$ plays the role of the cutoff function in the quantum gravity perturbation
theory expanded against a fixed background spacetime at all orders.
\subsection{The first case}
The bilocal operator is an entire function of order higher than or equal to 1/2, and we will consider it initially as 1/2,  so that $\mathcal{F}(\Box_x)=e^{\frac{1}{2}\ell\ \Box_x^{1/2}}$. Then, the modified density of energy is \cite{Nicolini:2012eu,2014jwq}
\begin{equation}\label{ModifiedDensity}
\tilde{\rho}({\bf x})=M\mathcal{F}^{2}(\Box_x)\delta({\bf x})=
\frac{M}{(2\pi)^2}\int d^2p\,e^{-\ell p}e^{i{\bf p\cdot x}}.
\end{equation}
In polar coordinates, we get
\begin{equation}
\tilde{\rho}(r)=\frac{M}{2\pi}\int_0^{\infty }dp\,p\, e^{-\ell p}J_0(p r) =\frac{ M\ell}{ 2\pi\left(r^2+\ell^2\right)^{3/2}}.
\end{equation}
This density profile is consistent with $\int_0^{\infty}\tilde{\rho}(r) 2\pi r dr=M$, as expected. Notice that this energy density is a decreasing function of the radial coordinate, with a maximum at the origin. This value could be of order of the Planck scales $M_p$ and $\ell_p$, {\it i.e} $\rho(0)=M_p/(2 \ell_p^3)$. As mentioned in the Introduction, a possible interpretation of this result is given in reference \cite{Rovelli:2014cta}, namely that a quantum phase is formed where the gravitational attraction is balanced by a very large quantum pressure.
Since the modified density of energy is now smoothed by the bilocal operator, we use the $00$--component
in Eq.(\ref{EME}) to obtain {the metric potential
\begin{equation} \label{funcionf}
f(r)=1-M+\frac{M\ell}{\sqrt{\ell^2+r^2}}-\Lambda r^2,
\end{equation}
where we considered $\kappa=1$, $\Lambda<0$.}% ( and redefined the mass $M/\pi\to M$\ ...).
In order to avoid singularities in the metric, the parameter $\ell$ is real and positive. Here the integration constant $1-M$ was chosen so as to yield a flat solution at the origin. Notice that
the solution is AdS-like for large $r$. Thus, we have
\begin{equation}\label{Sol1}
ds^2=-\left[1- M\left(1-\frac{\ell}{\sqrt{\ell^2+r^2}}\right)-\Lambda r^2\right]dt^2 +\frac{dr^2}{1- M\left(1-\frac{\ell}{\sqrt{\ell^2+r^2}}\right)-\Lambda r^2}+r^2d\phi^2.
\end{equation}
Notice that this solution retrieves the BTZ features for large $r$ and large masses, provided $\Lambda<0$. The values of the Ricci and Kretschmann invariants of curvature are
\begin{equation}
R = \frac{3 \ell^3 M}{\left(\ell^2+r^2\right)^{5/2}}+6\Lambda,
\end{equation}
{\begin{equation}
  K=  \frac{M^2\ell^2 (\ell^2-2r^2)^2}{(\ell^2+r^2)^5}+2 \left [ \frac{M\ell}{(\ell^2+r^2)^{3/2}} +2 \Lambda \right]^2.
\end{equation}
Both invariants are free of singularities everywhere making evident the regular nature of this solution.
For convenience, we define the real constant $b$ such that $\Lambda=-1/b^2$.
The potential function given by Eq.\eqref{funcionf} near the origin behaves as:
\begin{equation} \label{near0sol1}
  f|_{r<<1} \approx 1 - \left (  \frac{M}{2 \ell^2} -\frac{1}{b^2}\right )r^2 = 1 - \Lambda_{eff}r^2
\end{equation}
where $\Lambda_{eff}$ represents an effective cosmological constant. As a consequence, for $\frac{M}{2 \ell^2} >\frac{1}{b^2} \to \Lambda_{eff}>0$, the solution behaves as a de-Sitter spacetime near the origin. Thus, due to the quantum nature of the energy density, which could be of the order of the Planck scale around the origin, the positive effective cosmological constant could result in a repulsive force \cite{DeLorenzo:2014pta}. It is worth noting that although $\Lambda_{eff}>0$ near the origin the value of the classical cosmological constant $\Lambda=-1/b^2$ is negative and so a cosmological horizon for the solution is still avoided.
On the other hand, for $\frac{M}{2 \ell^2} =\frac{1}{b^2} \to \Lambda_{eff}=0$, the spacetime has a flat core and for $\frac{M}{2 \ell^2} <\frac{1}{b^2} \to \Lambda_{eff}<0$, the solution has an AdS core. It is worth stressing that in both cases the solution is regular. The search for an explanation connecting these possibilities with the nature of a Planckian energy density near the origin is of deep physical interest to be explored in future work.

\subsection{The second case}

{We will now work with the first order bilocal operator {\it i.e.} $\mathcal{F}^{-2}(p^2)=e^{-\ell^2 p^2}$.
The modified energy density, obtained from Eq. (\ref{ModifiedDensity}) is given by
\begin{equation}
\tilde{\rho}(r)= M\ \frac{ e^{-\frac{r^2}{4 \ell^2}}}{4\pi\ell^2},
\end{equation}
showing} that the bilocal operator diffuses the matter-energy according to a Gaussian distribution around the origin.
The profile is thus decreasing with respect to the radial coordinate and has its maximum value at the origin. This maximum value could be of the order of the Planck scale, {\it i.e.} $\tilde{\rho}(0)=M_p/(4 \ell_p^2)$, and so be associated with quantum effects as above mentioned.

The metric potential can be obtained using Eq. (\ref{EME}), yielding the line element
\begin{equation}\label{Sol2}
ds^2=-\left[1-M\left(1-e^{-\frac{r^2}{4 \ell^2}}\right)-\Lambda r^2\right]dt^2 +\frac{dr^2}{1-M\left(1-e^{-\frac{r^2}{4 \ell^2}}\right)-\Lambda r^2}+r^2d\phi^2.
\end{equation}
The integration constant is the same as in the previous case, chosen in order
to set the flatness of the solution at the origin.
For $\Lambda<0$ note again the convergence to the BTZ solution for large $r$ and $M$.

The values of the Ricci and Kretschmann curvature invariants are
\begin{equation}
  R=  \frac{M}{4 \ell^4}(6 \ell^2-r^2) \exp \left (\frac{-r^2}{4\ell^2}\right) + 6 \Lambda
\end{equation}
\begin{equation}
  K=  \left [ \frac{M}{4 \ell^4}(2 \ell^2-r^2) \exp \left (\frac{-r^2}{4\ell^2}\right) + 2 \Lambda \right ]^2 + 2 \left [ \frac{M}{2 \ell^2}\exp \left (\frac{-r^2}{4\ell^2}\right) + 2 \Lambda \right ]^2
\end{equation}

So, it is easy to check that the solution is free of singularities everywhere.
Again, the solution near the origin behaves as in eq. \ref{near0sol1}
%\begin{equation} \label{near0sol2}\nonumber
$  f|_{r<<1} \approx 1 - \left (  \frac{M}{4 \ell^2} -\frac{1}{b^2}\right )r^2 = 1 - \Lambda_{eff}r^2.$
%\end{equation}
As in the previous case, the value of the effective cosmological constant can be negative, positive, or zero and the considerations are the same as before.

\subsection{Remarks on the two cases}
{The profile energy densities derived from the bilocal operator, besides being finite at the origin also yield finite values after integration in the whole space $E=\int_0^{\infty} \tilde{\rho}(r)2\pi r dr = M < \infty$.
Following \cite{Aoki:2020prb} this value corresponds to the total energy computed with the conserved charge in both cases.
The source is said to have a quasi-localized mass/energy because it is so required by non-locality \cite{Estrada}.} Inversely, a local theory such as non-linear electrodynamics in 2+1 cannot be the cause of a quasi-localized material source \cite{Alana}. This seems to be a general result. Note both profile densities follow the requirements mentioned in reference \cite{Aros:2019quj} for quasi-localized mass/energy,  namely to be continuous and differentiable decreasing functions with a finite maximum at the origin and tending to zero at infinity.

In Fig. \ref{ModifiedDensity} we display the behavior of the $M$ parameter for the first solution, consistent with $f(r,M)=0$ which is in order to determine the number of existing horizons.
Note that there exists a critical value of $M=M_{cri}$ such that for $M < M_{cri}$ there would be no horizons. For $M=M_{cri}$ the black hole is said to be extremal, i.e. the inner and the event (external, indicated by $r_h$) horizons coincide. For $M>M_{cri}$ the two horizons are well separated.
It is easy to check that the second solution has the same general behavior.

\begin{figure}
\centering
 	\includegraphics[width=0.65\textwidth]{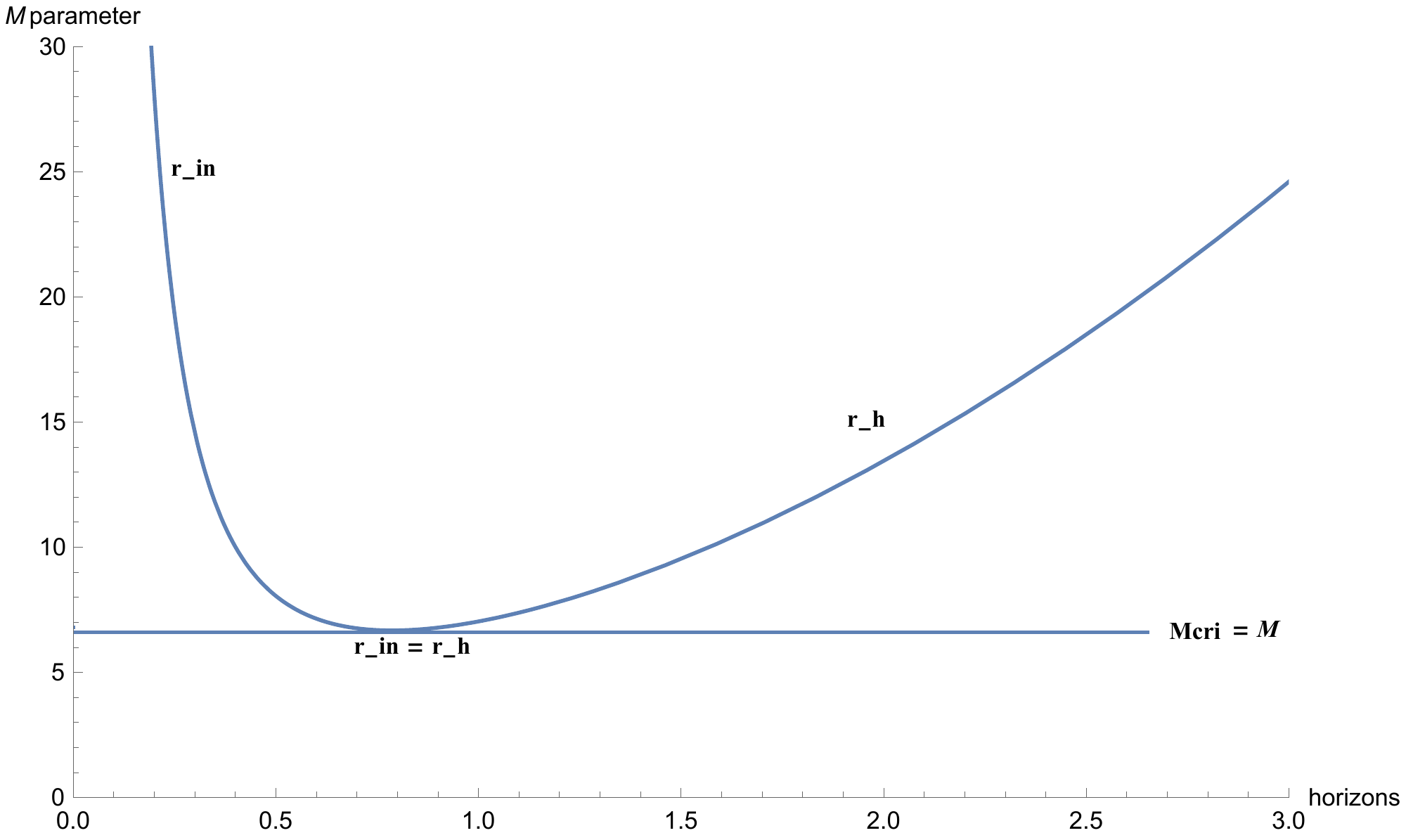}
 	\caption{Behavior of the $M$ parameter for $\ell=0.7$ and $\Lambda=-2.0$.}
 	\label{Mparametro}
 \end{figure}

In Fig. \ref{f(r)graph}, we show the metric potential with the two horizons for the black hole solutions just found. Notice that the horizons of the second solution, Eq.(\ref{Sol2}), are further from the center than the ones associated with the first solution, Eq. (\ref{Sol1}).

\begin{figure}
\centering
 	\includegraphics[width=0.55\textwidth]{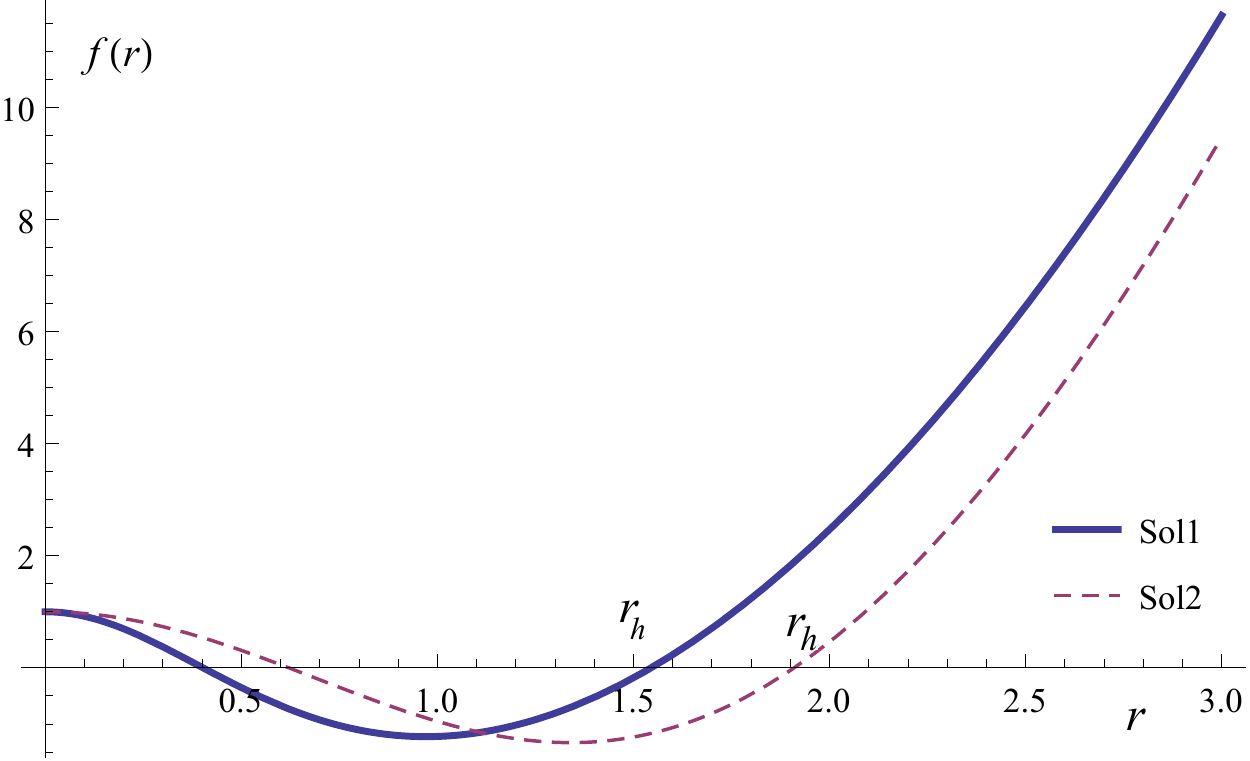}
 	\caption{Potential metric of the found bilocal black hole solutions (Sol1 in Eq. (\ref{Sol1}) and Sol2 in Eq. (\ref{Sol2})), for $M=10$, $\ell=0.7$, and $\Lambda=-2.0$.}
 	\label{f(r)graph}
 \end{figure}

 \section{Energy conditions}

We now analyze the energy conditions for the source of the modified black hole solution.
Since the source is scattered throughout space due to the nonlocal behavior of gravity, we must find the regions where the Null Energy Conditions (NEC, $\rho+p_i\geq 0$), Weak Energy Conditions (WEC, $\rho\geq 0$, $\rho+p_i\geq 0$), Strong Energy Conditions (SEC, $\rho+p_i\geq 0$, $\rho+\sum p_i\geq 0$), and Dominant Energy Conditions (DEC, $\rho\geq 0$, $-\rho\geq p_i\geq\rho$) are satisfied. The index $i$ is for the polar coordinates, $(1,2)\equiv (r,\phi)$. The first thing to observe is that in Eq. \eqref{Sol1} just WEC (and therefore NEC) are globally satisfied, since $\rho=-p_1$, and $\rho$, $\rho+p_2$ are always positive.
SEC will be satisfied in the region $r\geq \ell/\sqrt{2}$ ($\forall M\geq 0$) as well as DEC are fulfilled in $r\leq\ell\sqrt{2}$. Thus, both conditions are satisfied jointly in the range $\ell/\sqrt{2}\leq r\leq\ell\sqrt{2}$, {\it i.e.} only in this region energy cannot be exotic. In Fig. \ref{energy_pressure} we depict the combinations of energy density and pressures in order to visualize these conditions.
\begin{figure}[h!]
    \centering
    \begin{minipage}{0.7\textwidth}  \nonumber
        \centering
        \includegraphics[width=0.95\textwidth]{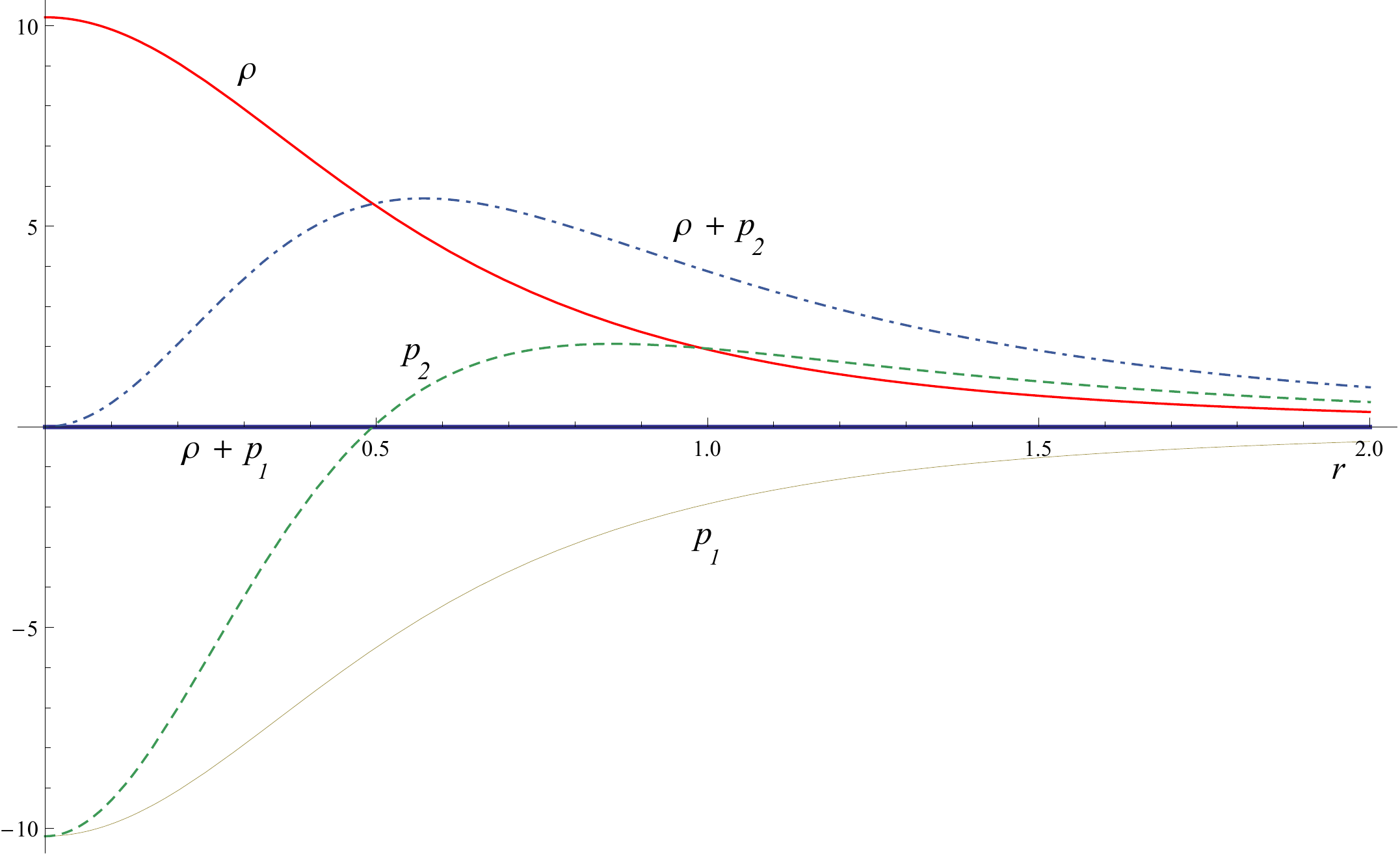}
    \end{minipage}\hfill
        \caption{Radial behavior of the energy density, pressures, and their combinations for the black hole solution given by Eq. \eqref{Sol1}. We have considered $\ell=0.7$ and $M=10$, in Planck units. The plots for solution Eq. (\ref{Sol2}) are quite similar.}
    \label{energy_pressure}
\end{figure}

The examination of the energy conditions satisfied by the black hole solution Eq. (\ref{Sol2}) results as in the first case: WEC (and NEC) are globally satisfied, SEC for $r\geq \ell\sqrt{2}$, and DEC for $r\leq 2\ell$. Therefore, all energy conditions are fulfilled only in the region $\ell\sqrt{2}\leq r\leq 2\ell$ so the matter can be assumed to be non-exotic. The plot is quite similar to the previous case and it is rather unnecessary to include it here.

\section{Thermodynamics}

Let us first consider the solution given by Eqs. (\ref{Sol1}) and (\ref{Sol2}) .
In these cases the Hawking temperatures, $T_H=f'(r_h)/4\pi$, are given by
 \begin{eqnarray}
 T_H^{sol1}&=&-\frac{\Lambda  r_h}{2 \pi } \left[1-\frac{\ell \left(\Lambda  r_h^2-1\right)}{2  \Lambda  \left(\ell^2+r_h^2\right) \left(\sqrt{\ell^2+r_h^2}-\ell\right)}\right],\nonumber\\
T_H^{sol2}&=&-\frac{\Lambda  r_h}{2 \pi } \left[1-\frac{\left(\Lambda  r_h^2-1\right)}{2 \Lambda\ell^2 \left(e^{\frac{r_h^2}{4 \ell^2}}-1\right)}\right],
 \end{eqnarray}
where the factor outside the brackets is the Hawking temperature for the BTZ black hole, provided $\Lambda<0$. In general, the behavior of these temperatures is asymptotically similar to the BTZ black hole one. For large $r_h$ are identical and for small $r_h$  the temperature decreases together with the horizon radius (the thermal capacity is positive). However, in the present cases, there exist phase transitions of zeroth order in which the temperatures vanish and the black hole evaporation halts at finite horizon radii. Hence, remnant masses appear as a result of phase transitions not present in the usual BTZ black hole. Fig. \ref{Temp} illustrates these features.
 \begin{figure}[h!]
    \centering
    \includegraphics[scale=0.8]{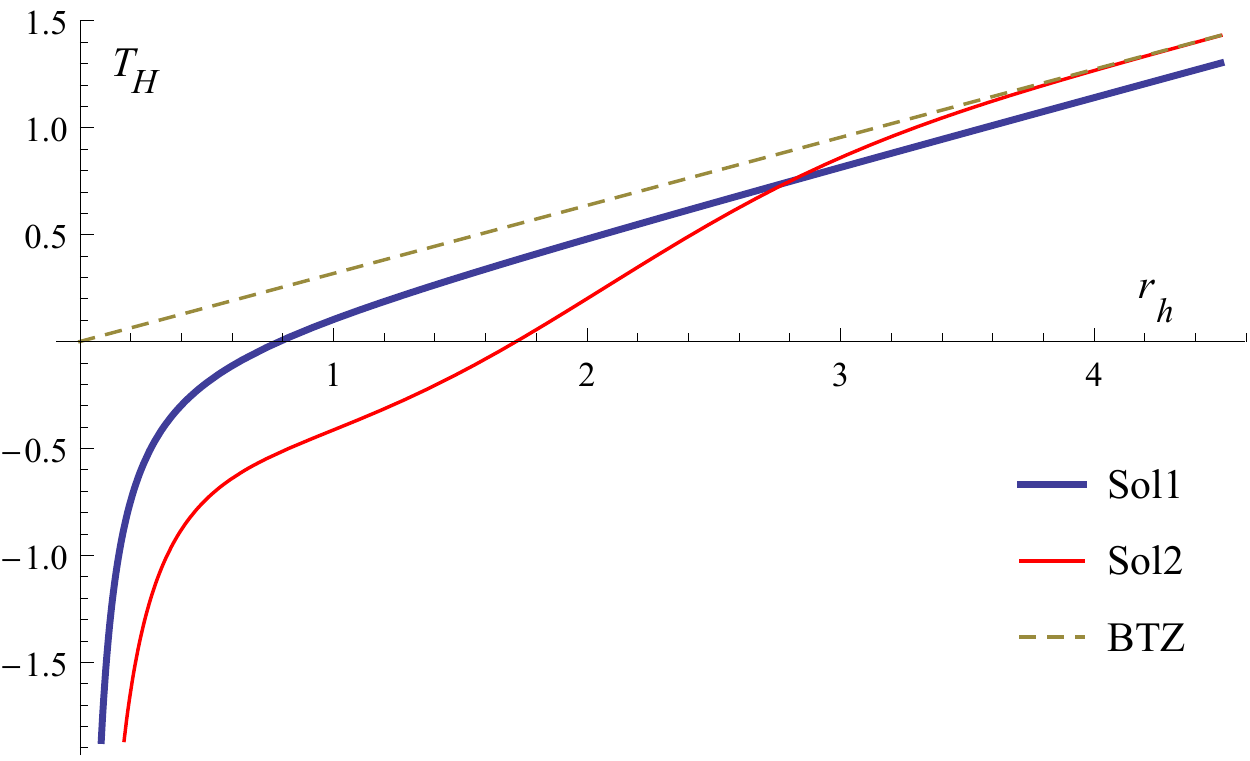}
    \caption{Hawking temperature as a function of the horizon radius, for the two analyzed solutions. Parameter settings: $\Lambda=-2.0$ and $\ell=0.7$.}
    \label{Temp}
\end{figure}

The critical horizon radius, $r_h^{*}$, at which the transition occurs can be analytically calculated for the second black hole solution; it yields
  \begin{equation}
r_h^{*}\geq \Lambda^{-1/2}\left[1-2 \Lambda  \ell^2-4 \Lambda  \ell^2 W_{-1}\left(-\frac{1}{2} e^{\frac{1}{4 \Lambda  \ell^2}-\frac{1}{2}}\right)\right]^{1/2},
\end{equation}
where $W_{-1}(x)$ is the second branch of the Lambert function.

It is well known the fact that the first law of thermodynamics must be modified for regular black holes due to the presence of matter fields in the energy-momentum tensor \cite{Estrada,Alana,Ma:2014qma}. This is because the usual version of the first law leads to incorrect values. For the sake of simplicity, we will follow the approach of reference \cite{Estrada}. This is based on the constraints in the space of parameters such that $f(r_h,M,p)=0$ and $\delta f(r_h,M,p)=0$. The thermodynamics pressure $p$ is associated to the cosmological constant such that $p=- \dfrac{\Lambda}{8\pi}$.

It is worth noting that in the case of solution \eqref{Sol1} as well as solution \eqref{Sol2}, the metric potential can be written as $f=1-m(r)- \Lambda r^2$. So, following \cite{Estrada} the first law of thermodynamics reads
\begin{equation} \label{duTdS}
du=TdS+ V dp  ,
\end{equation}
where $du$ corresponds to a local definition of the variation of the energy at the horizon, such that $du=\frac{\partial m}{\partial M} dM$. For this reason, its value does not coincide with the total energy $E$ \cite{Aoki:2020prb}. The entropy and black hole volume are respectively $S=\frac{\pi r_h}{2}$ and $V=\pi r_h^2$.

The heat capacity at constant pressure is given by
\begin{eqnarray}
\nonumber C_P&=&T\left(\frac{\partial S}{\partial T}\right)_P=T\left(\frac{\partial S/\partial r_h}{\partial T/\partial r_h}\right)_P=T\left(\frac{\pi/2}{\partial T/\partial r_h}\right)_P.
\end{eqnarray}
For the present black holes, because the factor $\partial T/\partial r_h$ is always positive, it is possible to show that the behavior of this quantity follows exactly the Hawking temperature, as shown in Fig. \ref{Temp}. Thus, $C_P$ is positive for $r_h > r_h^{*}$, indicating that the black holes are locally stable while they are irradiating. This behavior also differs from the vacuum BTZ black hole, where the heat capacity is always positive and depends linearly on the horizon radius.% $C_p \sim r_h$.
\begin{figure}[h!]
    \centering
    \includegraphics[scale=0.8]{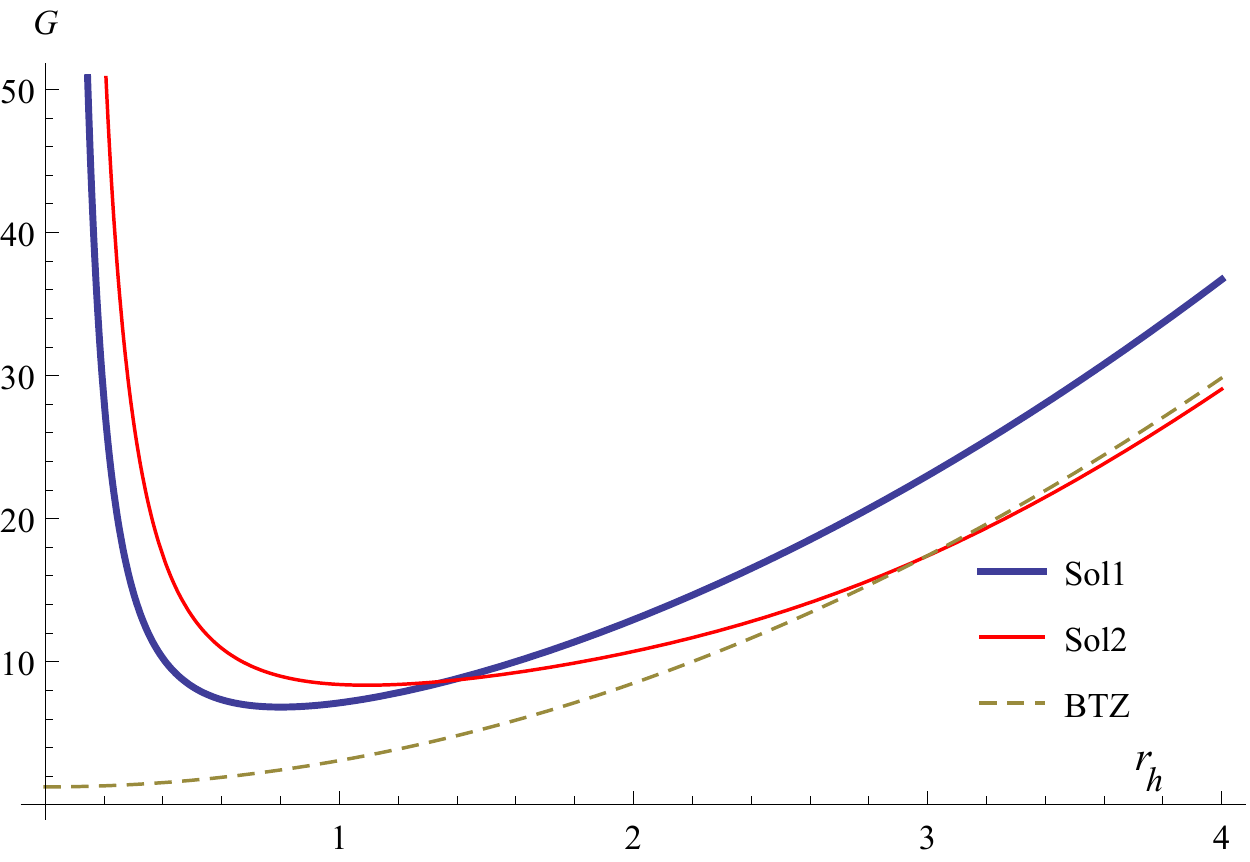}
    \caption{Gibs free energy of the bilocal black hole solutions as a function of the horizon radius. We have considered $\ell=0.7$ and $\Lambda=-2.0$.}
    \label{thermo1}
\end{figure}

We now analyze the Gibbs free energy, given by
\begin{eqnarray}
\nonumber G&=&M-TS+PV.
\end{eqnarray}
In Fig. \ref{thermo1} we depict this quantity as a function of the horizon radius. Notice that none of the black hole solutions are globally stable, since $G>0$ for all $r_h$. Notice also that the Gibbs free energy tends to BTZ's one for large horizons. Such a behavior is quite similar to the mass (internal energy) shape, as shown in Fig. \ref{Mparametro}.

\section{Geodesic circular orbits}

The geodesics of a particle in orbit around a static black hole are given by
\begin{equation}
\dot{r}^2=  \mathcal{E}^2 - f(r)\left(\frac{\lambda^2}{r^2}-\epsilon\right),
\end{equation}
where $\dot{r}$ is the derivative with respect to the proper time, $\mathcal{E}$ is the particle's energy per mass unit, $\lambda$ is the angular momentum per mass unit, $\epsilon=-1$ for a massive particle and $\epsilon=0$ for light.
The effective potential is defined as
\beq
V_r=f(r)\left(\frac{\lambda^2}{r^2}-\epsilon\right).
\eeq\label{effectpot}
\begin{figure}[htb!]
    \centering
            \includegraphics[width=0.6\textwidth]{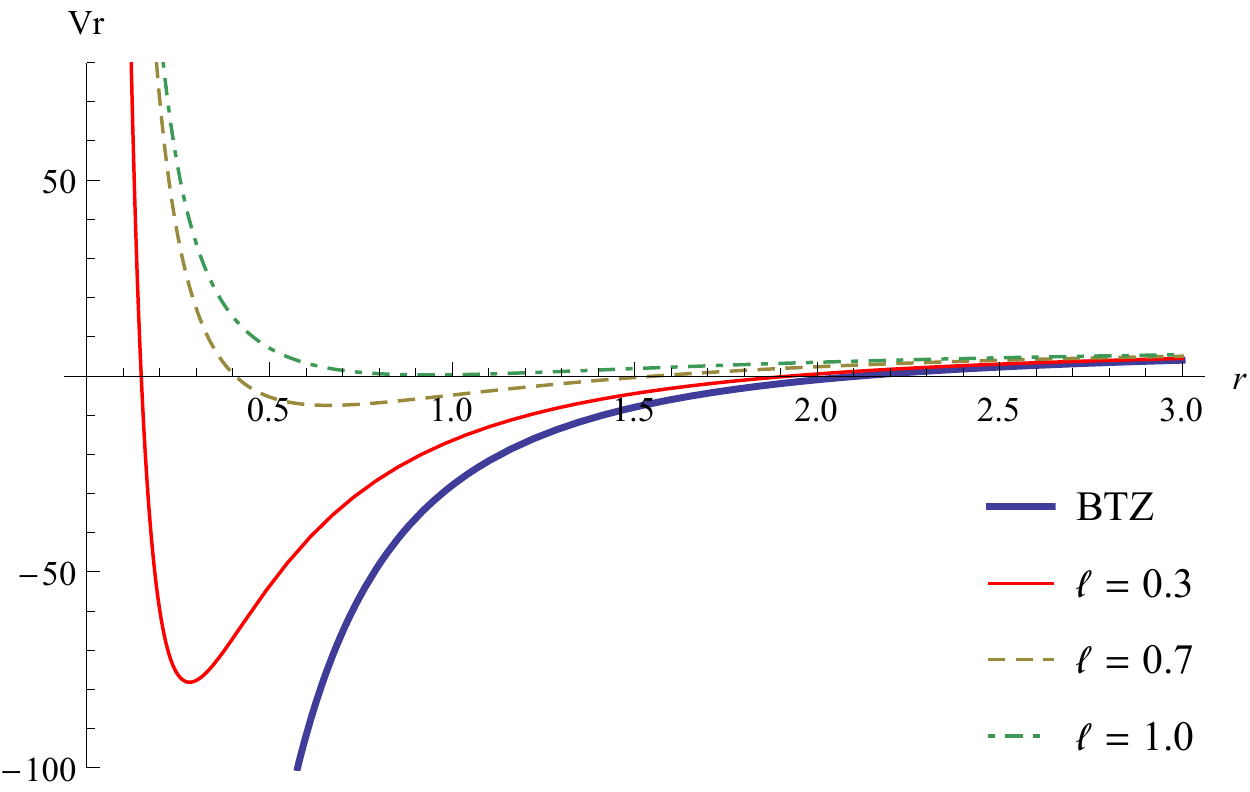}
        \caption{Effective potential as a function of the radial coordinate, $r$, for {\it photons} around the first solution of the 3-dimensional bilocal black hole. The parameter settings are $\Lambda=-2.0$, $M=10.0$, and $\lambda=2.0$, for some values of $\ell$, in Planck units. }
    \label{Phorb}
\end{figure}
\begin{figure}[htb!]
    \centering
            \includegraphics[width=0.63\textwidth]{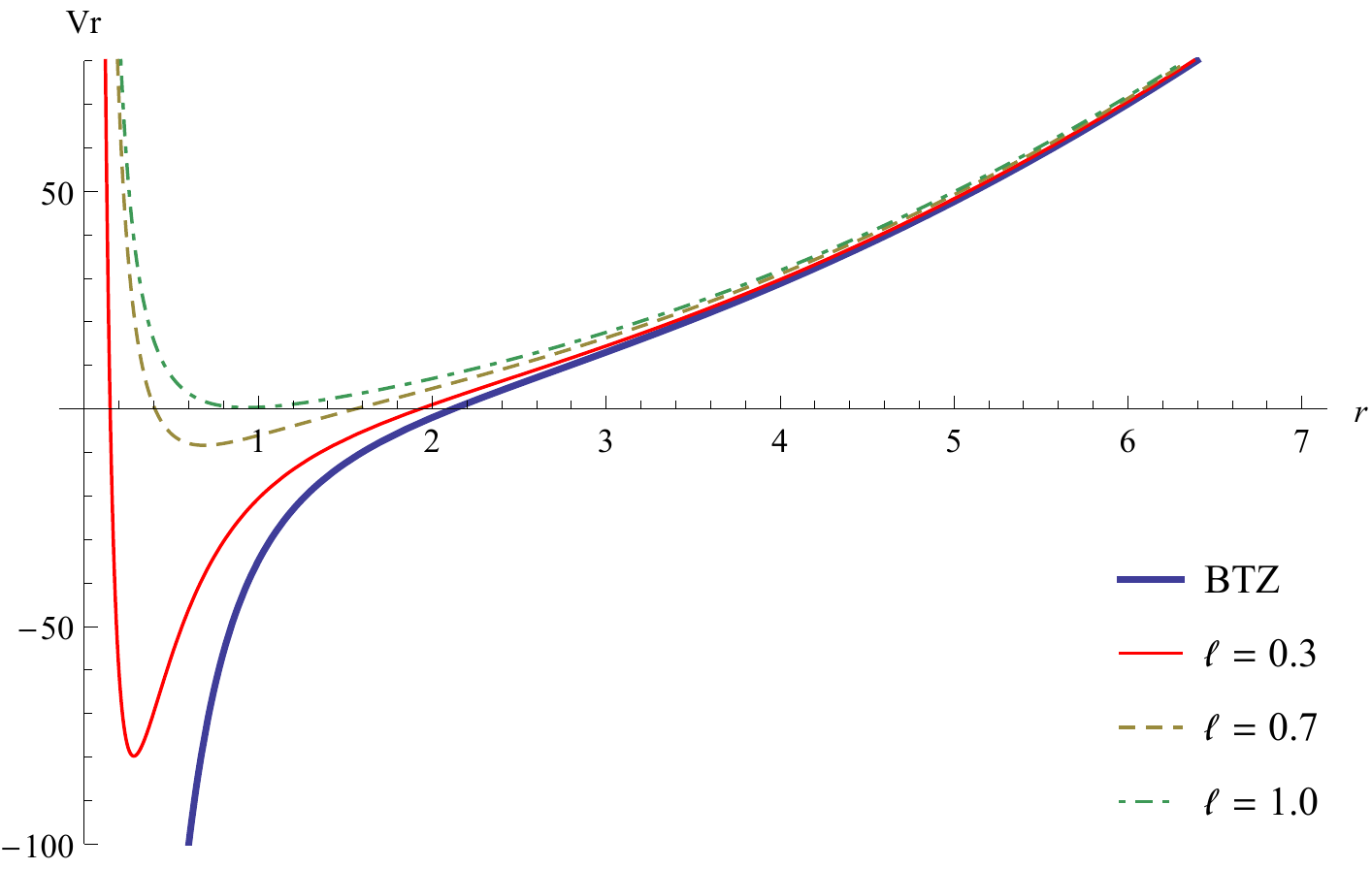}
        \caption{Effective potential as a function of the radial coordinate, $r$, for {\it massive} particles around the first solution of the 3-dimensional bilocal black hole. The parameter settings are $\Lambda=-2.0$, $M=10.0$, and $\lambda=2.0$, for some values of $\ell$, in Planck units.}
    \label{Morb}
\end{figure}
The circular geodesics occur at the points $r_c$ satisfying $\dot{r_c}=0$ and $V'_r (r_c)=0$. Fig \ref{Phorb} accounts for the effective potential in the case of photons, considering solution 1. This potential asymptotically tends to $-\Lambda \lambda^2$.  Notice the existence of circular photon orbits for $\ell\neq 0$, which are stable since $V_r$ has a local minimum. These orbits are localized between the inner horizon and the external one. Such a feature is completely different from the BTZ solution, for which photon orbits do not occur as we can see in the graphic.
The shape of the effective potential associated with solution 2 is very similar and we shall not exhibit it. However, for this solution, we can calculate the radius of the circular photon orbits analytically, finding
\begin{equation}
r_c^{sol2}=2 \ell \sqrt{-W_{-1}\left(\frac{1-M}{e M}\right)-1},
\end{equation}
where $W_{-1}(x)$ is Lambert's function. It will be a real number if $M>1$, as expected. We can also note its independence from the cosmological constant, which is also valid for the first solution.

Regarding massive particles (see Fig. \ref{Morb}) the effective potential diverges asymptotically but has a local minimum as well. Thus, stable circular geodesics placed between the horizons are also possible around the bilocal black holes. Fig. \ref{Morb} illustrates these features for solution 1. The black hole solution 2 has quite the same behavior so we shall not depict it here. 

\section{Conclusions}

In this paper, we have obtained regular black hole solutions for a bilocal gravity action in 2+1 dimensions with a negative cosmological constant. We have considered two different bilocal operators which depend on a fundamental length, $\ell$. The solutions found can be thought of as being generated by effective quasi-localized material sources. {Following \cite{Aoki:2020prb}, the value of the total energy, computed with the conserved charge, is $E = M$ in both cases}. We have shown that the solutions are free of singularities, as expected, and present up to two horizon radii. These solutions converge to the classical GR-based uncharged BTZ solution for large distances and large BH masses $M$.
Both energy density profiles have a finite maximum at the origin which could be of the order of the Planck scale. This could be associated with the formation of a quantum gravitational phase, where the gravitational attraction is balanced by a very large quantum pressure \cite{Rovelli:2014cta}. Near the origin, the solution can have a dS, AdS, or flat core depending on the relation among the parameters. To the de Sitter core corresponds a positive effective cosmological constant which could be viewed as a repulsive force due to quantum effects \cite{DeLorenzo:2014pta}. This is a consequence of the nature and magnitude of the energy density, which could be of the order of the Planck scale near the origin. It is worth noticing that although the effective cosmological constant near the origin is positive, the value of the classical naked $\Lambda=-1/b^2$ is still negative. So,  the advent of a cosmological horizon is avoided.

We have analyzed the energy conditions on the effective source of the new regular black hole solutions. As a result, the supporting energy is necessarily non-exotic only in a small range of the radial coordinate where all the energy conditions are satisfied (WEC, SEC, and DEC).
We have calculated some relevant thermodynamic quantities for the solutions obtained (Hawking temperature, BH volume, heat capacity at constant pressure, and Gibbs free energy) and analyzed the effect of a nonlocal source as compared with the usual BTZ case. On the one hand, the analysis of the Gibbs free energy has shown that the solution (\ref{Sol1}) is globally unstable ($G>0$) for all $r_h$, as in the BTZ case. On the other,
for small values of the horizon radius, $T_H$ is negatively divergent but a finite size remnant can be defined at a critical $r_h^{*}$, where the Hawking temperature vanishes. The mass remnant for solution Eq.(\ref{Sol1}) is smaller than the remnant related to solution Eq.(\ref{Sol2}). The heat capacity at constant pressure revealed that both the bilocal black holes are locally stable for $r_h > r_h^{*}$ ({\it i.e.}, $C_P>0$ along with $T_H\geq 0$).
At this point the heat capacity sign changes from negative to positive, indicating that the black holes are locally stable while they are irradiating. In conclusion, both $G$ and $T_H$ present crucial differences with the BTZ black hole thermodynamics for small horizon radii, where quantum effects become relevant.

Finally, we have analyzed and remarked on the existence of stable circular geodesics for both massless and massive particles around the three-dimensional bilocal black holes. This is another feature that is not observed in the BTZ solution.

\acknowledgments{The authors would like to thank Conselho Nacional de Desenvolvimento Cient\'{i}fico e Tecnol\'{o}gico (CNPq) for the partial financial support. Milko Estrada is funded by the FONDECYT Iniciaci\'on Grant
11230247.}

\section*{Data availability statement}

Data sharing is not applicable to this article as no datasets were generated or analyzed during the current study.

\end{document}